\documentclass[showpacs,aps,amssymb,pra,amsmath]{revtex4}
\usepackage{float}
\usepackage{epsfig}
\usepackage{bm}
\usepackage{amsmath,graphicx}
\usepackage{subfig}
\usepackage{setspace}
\usepackage{color}
\renewcommand{\vec}{\mathbf}

\newcommand{\beq}{\begin{equation}}
\newcommand{\eeq}{\end{equation}}
\newcommand{\bea}{\begin{eqnarray}}
\newcommand{\eea}{\end{eqnarray}}

\def\br{{\bf r}}

\def\bF{{\bf F}}
\newcommand{\Cdd}{C_{{dd}}}
\newcommand{\bCdd}{{C}_{{dd}}}

\begin{document}
\title{Collective excitations of a harmonically trapped, two-dimensional, spin-polarized dipolar Fermi gas in the hydrodynamic regime}

\author{B. P. van Zyl}
\affiliation{Department of Physics, St. Francis Xavier University, Antigonish, Nova Scotia, Canada
B2G 2W5}
\author{E. Zaremba}
\affiliation{Department of Physics, Astronomy and Engineering Physics, Queen's University, Kingston, Ontario, Canada K7L 3N6}
\author{and J. Towers}
\affiliation{Jack Dodd Centre for Quantum Technology, Department of Physics, University of Otago, Dunedin, New Zealand}


\date{\today}
\begin{abstract}
The  collective excitations of a zero-temperature, spin-polarized, harmonically trapped, two-dimensional dipolar Fermi gas are examined within the Thomas-Fermi von Weizs\"acker
hydrodynamic theory.   We focus on repulsive interactions, and investigate the dependence of the excitation frequencies on the strength of the dipolar interaction and particle number.
We find that the mode spectrum can be classified according to bulk modes, whose frequencies are shifted upward as the interaction strength is increased, and an infinite ladder of surface modes, whose frequencies are {\em independent} of the interactions in the large particle limit.  We argue quite generally that it is the {\em local} character of
the two-dimensional energy density  which
is responsible for the insensitivity of surface excitations to the dipolar interaction strength, and not the precise form of the equation of state.    
This property will not be found for the collective excitations of harmonically trapped, dipolar Fermi gases in one and three dimensions, where the energy density is manifestly nonlocal.
\end{abstract}
\maketitle
\section{Introduction}
In a recent paper~\cite{vanzyl_pisarski}, we presented a fully self-consistent, density-functional theory~\cite{DFT} for the description of the equilibrium properties
of a harmonically trapped, two-dimensional (2D) spin-polarized dipolar Fermi gas (dFG) at zero-temperature $(T=0)$.  For  dipoles oriented perpendicular to the 2D $x$-$y$ plane, the dipole-dipole interaction is isotropic, and strictly repulsive.

At the heart of the formulation is the Thomas-Fermi  von Weizs\"acker (TFvW)
energy functional, {\it viz.,}
\bea
\label{ETFvW}
E[n] &=& \int d^2r \left[\frac{1}{2}C_K [n(\br)]^2 + 
 \frac{2}{5}C_{dd} [n(\br)]^{5/2}\right] + \int d^2r~ \lambda_{\rm vW}\frac{\hbar^2}{8M}\frac{|\nabla 
n(\br)|^2}{n(\br)}+ \int d^2r~v_{\rm ext}(\br) n(\br) \\
&\equiv& \int d^2 r ~\varepsilon_{\rm loc}[n] + \int d^2r ~\varepsilon_{\rm vW}[n]+ \int d^2r~v_{\rm ext}(\br) n(\br)\nonumber~,
\eea
where $C_K \equiv 2\pi\hbar^2/M$, $C_{dd} \equiv (32/9\sqrt{\pi})\mu_0D^2$, and $D$ is the magnetic moment of the neutral atoms.  
The external potential is given by
$v_{\rm ext}(\br) = \frac{1}{2} M \omega_0^2 r^2$, where $\omega_0$ is the isotropic 2D trap frequency { and $M$ is the mass of an atom.}
The first term of the local energy density, $\varepsilon_{\rm loc}[n]$, corresponds to the noninteracting kinetic energy of a uniform 2D Fermi gas.   The second term 
in $\varepsilon_{\rm loc}[n]$ is associated with the  {\em total} dipole-dipole interaction energy in the Hartree-Fock approximation~\cite{DFT}, which in 2D can be accurately represented as a purely local
function of the spatial density, $n(\br)$~\cite{vanzyl_pisarski,fang}.  Finally, the von Weizs\"acker (vW) energy density, $\varepsilon_{\rm vW}[n]$, is included to account for the increase in the kinetic energy associated with 
the spatial inhomogeneity introduced by
the external trapping potential~\cite{vW}.  The parameter $\lambda_{\rm vW}$, is the so-called vW coefficient,
which takes the value $\lambda_{\rm vW}\simeq 0.02-0.04$ for particle numbers in the range $N \sim 10^2$-$10^6$~\cite{vanzyl_pisarski}.

Introducing the vW wavefunction, $\psi(\br) \equiv \sqrt{n(\br)}$, and performing the
variational minimization of Eq.~(\ref{ETFvW}) with respect to the 
density, gives 
\beq\label{SE_TFvW}
-\lambda_{\rm vW}\frac{\hbar^2}{2M} \nabla^2 \psi(\br) + v_{\rm 
eff}(\br) \psi(\br) = \mu \psi(\br)~,
\eeq
where $v_{\rm eff}(\br)$ is the effective one-body potential given by
\bea\label{veff_TFvW}
v_{\rm eff}(\br)
&=& \frac{d \varepsilon_{\rm loc}[n]}{d n} + \frac{1}{2}M\omega_0^2r^2\nonumber\\
&=& C_K\psi^2(\br) + C_{dd} \psi^3(\br) + \frac{1}{2}M\omega_0^2r^2~.
\eea
Since $v_{\rm eff}(\br)$
itself depends on $\psi(\br)$, the solution of Eq.~(\ref{SE_TFvW})
must be determined self-consistently. The ground state solution,
$\psi_0(\br)$, determines the self-consistent ground state density $n_0(\br) = \psi^2_0(\br)$ 
and the chemical potential, $\mu$, is fixed by the normalization condition
\beq\label{particles}
\int d^2r ~\psi^2_0(\br) = N~.
\eeq
Once the solution to Eq.~(\ref{SE_TFvW}) is obtained, we have a complete description of the equilibrium properties of the system.  The numerical scheme used to solve for the
self-consistent ground state density is outlined in  Ref.~\cite{vanzyl_pisarski}.

In this paper, we extend our earlier work~\cite{vanzyl_pisarski} to consider the collective excitations that are induced by driving
the system away from its equilibrium ground state density, $n_0(\br)$.  The main motivation of this study is the recent realization of a degenerate, spin-polarized gas of  $^{161}$Dy atoms~\cite{liu}.
In view of the reasonable expectation that the quasi-2D analogue of this experiment is possible, along with foreseeable studies of the collective excitations, 
a theoretical investigation of the collective modes is of interest.  
Although a variety of theoretical techniques have been used
to investigate the collective excitations of trapped dFGs in three~\cite{watch,adhikari,abad,lima,zhao,lima2,sogo,baranov}, and lower dimensional geometries~\cite{abed,babadi,baranov3,baranov2}, we feel that it is useful to present another, relatively unknown approach, namely, the TFvW hydrodynamic theory.

The TFvW approach  in fact has its roots outside of cold-atom physics. It has had a long, and successful history in the description of the  collective
excitations of degenerate electron gases in 3D~\cite{zaremba_tso,zaremba96} and lower dimensions~\cite{vanzyl1,vanzyl2,vanzyl3,vanzyl4,PhD}.  
Owing to the fact that the TFvW theory is not fundamentally linked 
to any particular form of the interparticle interactions,  there is no {\em a priori} reason to believe that it should not be equally effective for studying the dynamics
of a degenerate dFG, provided that the system is assumed to be in the hydrodynamic (HD) regime.  In addition, the TFvW is numerically easy to implement for any number of atoms, $N$, and has the virtue
of treating the dynamics of the system in a way which is consistent with the equilibrium properties.

The outline of the rest of the paper is as follows. 
In Sec.~\ref{CE}, the TFvW hydrodynamic theory is presented,  along with a numerical analysis of the collective modes of the system as the strength of the dipolar interaction, and particle number, are varied.  
In Sec.~\ref{analytical}, we provide analytical support for our numerical calculations,  and in Sec.~\ref{closing} we present our closing
remarks and conclusions.

\section{TFvW Hydrodynamics}\label{CE}
The essence of the TFvW hydrodynamic theory for the collective excitations~\cite{zaremba_tso} is to treat the system as a ``classical'' fluid obeying the usual continuity equation
\beq\label{cont}
\frac{\partial n}{\partial t} + \nabla\cdot(n {\bf v}) =0~,
\eeq
and the momentum equation
\bea\label{mom}
M\left [ \frac{\partial {\bf v}}{\partial t} + {\bf v}\cdot \nabla {\bf v}\right] = {\bf F}~,
\eea
where $\bF$ accounts for total force acting on the atoms,
\beq\label{fint}
\bF(\br,t) = -\nabla \left[ v_{\rm eff}(\br,t) - \lambda_{\rm vW}\frac{\hbar^2}{2M} \frac{\nabla^2\psi(\br,t)}{\psi(\br,t)}\right]~.
\eeq
Here, $v_{\rm eff}(\br,t)$ is defined by Eq.~(\ref{veff_TFvW}) with the replacement of $\psi(\br)$ by the dynamic wave function $\psi(\br,t)$.
Note that when the bracketed quantity in Eq.~\eqref{fint} is evaluated for the (static) ground state $\psi_0(\br)$, it is equal to a constant, the chemical potential $\mu$. Thus the total force vanishes for the equilibrium situation.

The collective modes of the system correspond to small-amplitude oscillations around 
the ground state distribution, $n_0(\br)$.
Introducing the density fluctuation $\delta n(\br,t) = n(\br,t) - n_0(\br)= 2\psi_0(\br)\delta \psi(\br,t)$ and linearizing the hydrodynamic equations in the fluctuating variables, we have
\beq\label{lin_cont}
\frac{\partial \delta n}{\partial t} + \nabla\cdot(n_0{\bf v}) = 0~,
\eeq
and
\beq\label{lin_mom}
\frac{\partial {\bf v}}{\partial t} = \frac{\delta \bF}{M}~,
\eeq
where the fluctuating force is given by
\beq\label{fluc_force}
\delta \bF(\br,t) = -\nabla\left [ \delta v_{\rm eff}(\br,t) - \lambda_{\rm vW}\frac{\hbar^2}{2M \psi_0}\nabla^2\delta \psi(\br,t) + \lambda_{\rm vW}\frac{\hbar^2}{2M} \frac{\nabla^2 \psi_0}{\psi_0^2} \delta \psi(\br,t)\right]~,
\eeq
with
\bea\label{fluc_veff}
\delta v_{\rm eff}(\br,t) 
&=& \left. \frac{d^2\varepsilon_{\rm loc}[n]}{dn^2}\right |_{n=n_0} \delta n(\br,t)\nonumber \\
&=& 2 C_K \psi_0 \delta\psi(\br,t) + 3 C_{dd} \psi_0^2 \delta \psi(\br,t)~.
\eea
Defining the ground state TFvW Hamiltonian
\beq\label{hhat}
\hat{h} \equiv -\lambda_{\rm vW} \frac{\hbar^2}{2M} \nabla^2 + v^{0}_{\rm eff}-\mu~,
\eeq
where $v^{0}_{\rm eff}$ is  the effective potential evaluated at $\psi_0$, we have
\beq
\delta \bF(\br,t) = -\nabla \left[ \delta v_{\rm eff}(\br,t) + \frac{1}{\psi_0} \hat{h}~ \delta \psi(\br,t)\right].
\eeq

Keeping in mind that we are considering an isotropic harmonic oscillator (HO) confinement potential, it is useful to
scale all energies and lengths by $\hbar\omega_0$ and
$a_{\rm ho} = \sqrt{\hbar/M\omega_0}$, respectively.  In particular, we define the dimensionless variables
\beq\label{dimless}
\bar{\psi} = a_{\rm ho}\psi,~~~\bar{C}_K = \frac{C_K}{\hbar\omega_0 a^2_{\rm ho}}=2\pi,~~~~\bar{C}_{dd}=\frac{2\pi}{a_{\rm ho}} \frac{C_{dd}}{C_K}~,~~~\bar{\omega} = \frac{\omega}{\omega_0}~.
\eeq
Unless it is needed for clarity, we will for simplicity drop the bar notation in the following.
Assuming a harmonic time-dependence, $e^{-i\omega t}$, for the fluctuating quantities, Eqs.~\eqref{lin_cont} and~\eqref{lin_mom} may be combined to yield 
\beq\label{dyn1}
-\omega^2 \delta n(\br)-\nabla \cdot [n_0(\br) \nabla f(\br)]=0~,
\eeq
where
\beq\label{f}
f(\br) \equiv \delta v_{\rm eff}(\br) + \frac{1}{\psi_0} \hat{h} \delta \psi(\br)~.
\eeq

Equation~\eqref{dyn1} may be cast in the form of a matrix eigenvalue problem by introducing the orthonormal basis defined by
\beq\label{orthobasis}
\hat{h} \phi_i(\br) = \varepsilon_i \phi_i(\br),
\eeq
with
\beq
\int d^2r ~\phi_i^*(\br)\phi_j(\br) = \delta_{ij}.
\eeq
Expanding the wave function fluctuation as
\beq\label{flucpsi}
\delta\psi(\br) = \sum_i c_i \phi_i(\br)~,~~~~i=1,...,p_{\rm max}~,
\eeq
the eigenvalue problem for the modes can be shown to take the form~\cite{zaremba_tso}
\beq\label{modes}
\omega^2 c_i = \frac{\varepsilon_i}{\lambda_{\rm vW}} \sum_j M_{ij} c_j~,
\eeq
where
\beq\label{Mmatrix}
M_{ij} \equiv \int d^2r~\phi^*_i(\br) [2C_K\psi_0^2(\br) + 3C_{dd} \psi_0^3(\br)+\varepsilon_j]\phi_j(\br)~.
\eeq
The collective mode frequencies are given by $\omega$, and the corresponding mode densities, $\delta n(\br)$, are obtained from Eq.~\eqref{flucpsi} using $\delta n(\br) = 2\psi_0(\br)\delta\psi(\br)$.
Since the basis set, $\{\phi_i(\br)\} $, includes $\psi_0$ as one of its elements, we see that $\int d^2r ~\delta n(\br) \propto c_0$. The requirement that the mode density integrates to zero implies $c_0 = 0$.
In the case of 2D isotropic HO confinement, the circular symmetry allows us to take  $\phi_i(\br) \to \phi_{nm}(r,\theta)=u_{mn}(r)e^{im\theta}$, such that
\beq\label{ortho2}
\int d^2r~\phi_{nm}^*(\br) \phi_{n'm'}(\br) = \delta_{nn'}\delta_{mm'}~,
\eeq
which implies that the modes for different $m$-values are decoupled. 
In our numerical calculations we have considered modes with angular momentum $m$ up to a maximum of 9 and a nodal index $n$ up to 4. For all of these modes and the ranges of $C_{dd}$ and $N$ considered, 
a value of $p_{\rm max} =2500$ was found to be sufficiently large to obtain convergence for the eigenfrequencies
and mode densities.

\subsection{Numerical results}\label{numerics}
While the TFvW theory can, in principle, be used to calculate the modes for any number of particles, the results are only meaningful for a trapped 
dFG  in the HD regime.  In Ref.~\cite{babadi}, it was shown that the following inequality,
\beq\label{testHD}
N \left(\frac{a_{dd}}{a_{\rm ho}}\right)^2 \gg 1~,
\eeq
is the condition determining whether the system is in the HD regime.  In Eq.~\eqref{testHD} we have introduced the dipolar length $a_{dd} \equiv \mu_0 D^2 M/(4 \pi \hbar^2).$ 
Taking, for example, a  value of $\omega_0 = 2\pi\times 1500$ Hz, gives $(a_{dd}/a_{\rm ho})^2 \approx 10^{-2}$ for $^{161}$Dy, and we are just in the HD regime for $N \gtrsim 10^3$.  As the
value of $\omega_0$ increases, the HD region can be reached for smaller values of $N$.
We will be interested primarily in particle numbers in the range of $N \sim 10^3 -10^5$, for which experiments on, {\it e.g.}  $^{161}$Dy, may sensibly  be compared with the TFvW theory.

\begin{figure}[ht]
\centering \scalebox{0.3}
{\includegraphics{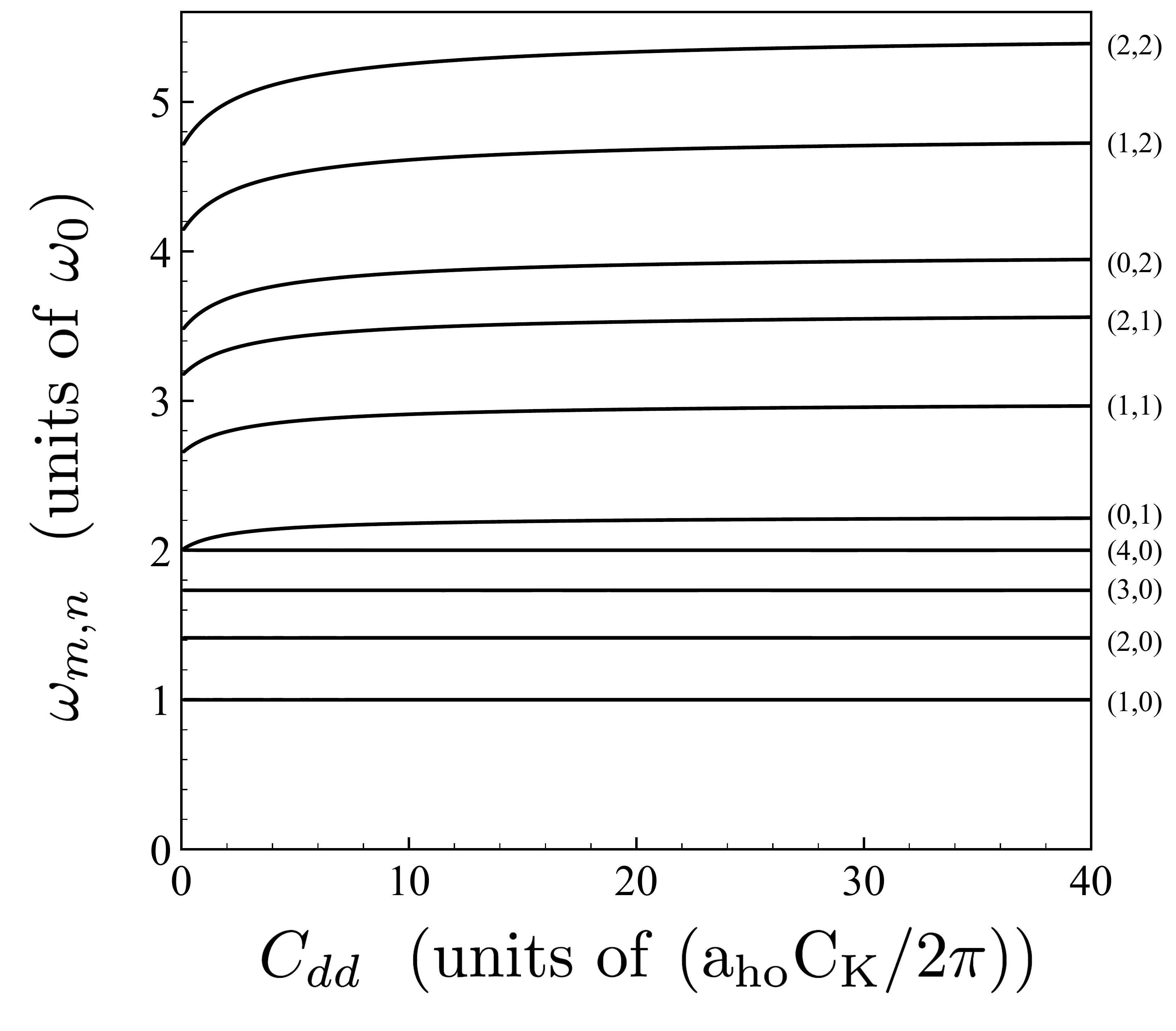}}
\caption{The collective excitation frequencies, $\omega_{m,n}$, for $N=10^3$ atoms.   ${C}_{dd}$ is dimensionless, as defined in Eq.~\eqref{dimless}, and the curves are labeled by  $(m,n)$.  
Note that for this particle number, the $\omega_{m,0}$ modes appear to be independent of the interaction strength on the scale of the plot. }
\label{fig1}
\end{figure}

\begin{figure}[ht]
\centering \scalebox{0.68}
{\includegraphics{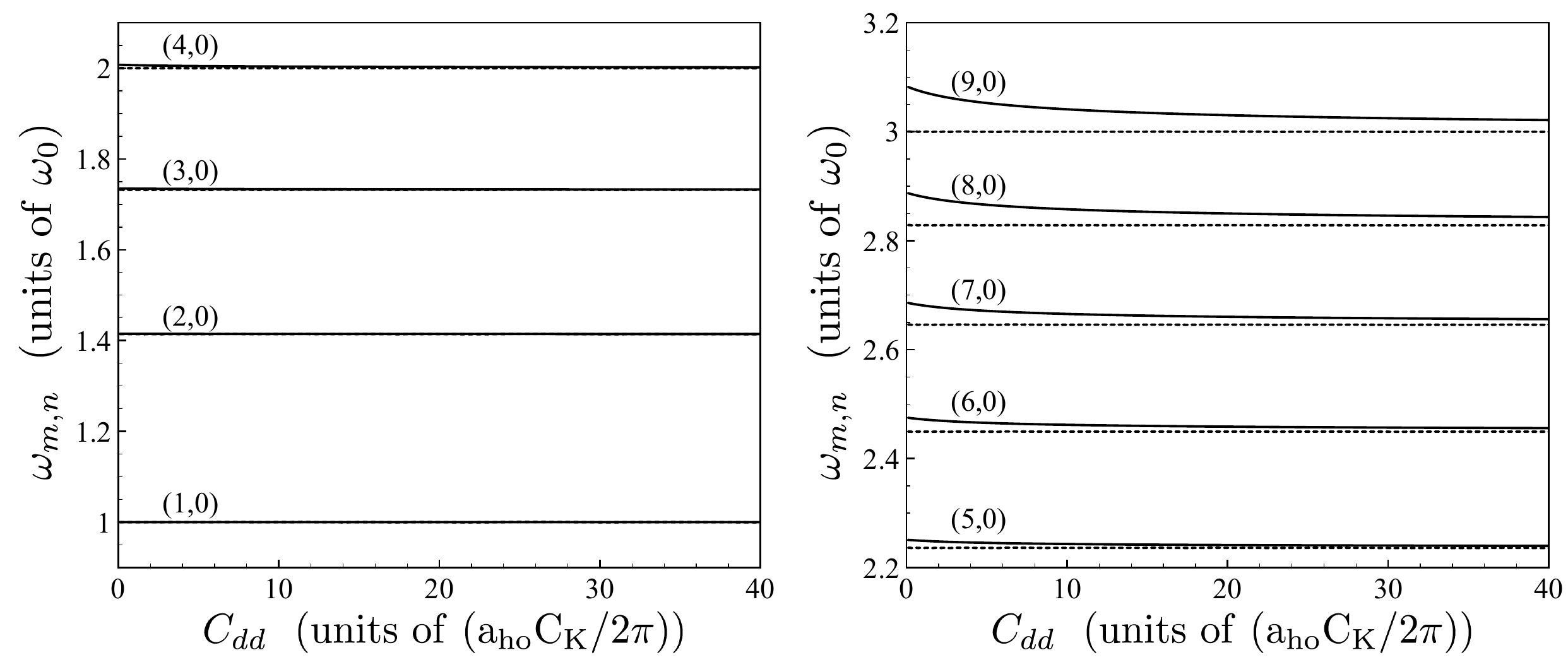}}
\caption{The  collective excitation frequencies, $\omega_{m,0}$, for $N=10^2$ (solid curves) and $N=10^5$ (dashed curves) atoms. This figure illustrates that for large particle numbers, the $\omega_{m,0}$
modes become independent of the interaction strength. As shown in Eq.~\eqref{w_m}, the dashed curves have the frequency $\omega_{m,0} = \sqrt{m}$.}
\label{fig2}
\end{figure}

\begin{figure}[ht]
\centering \scalebox{0.3}
{\includegraphics{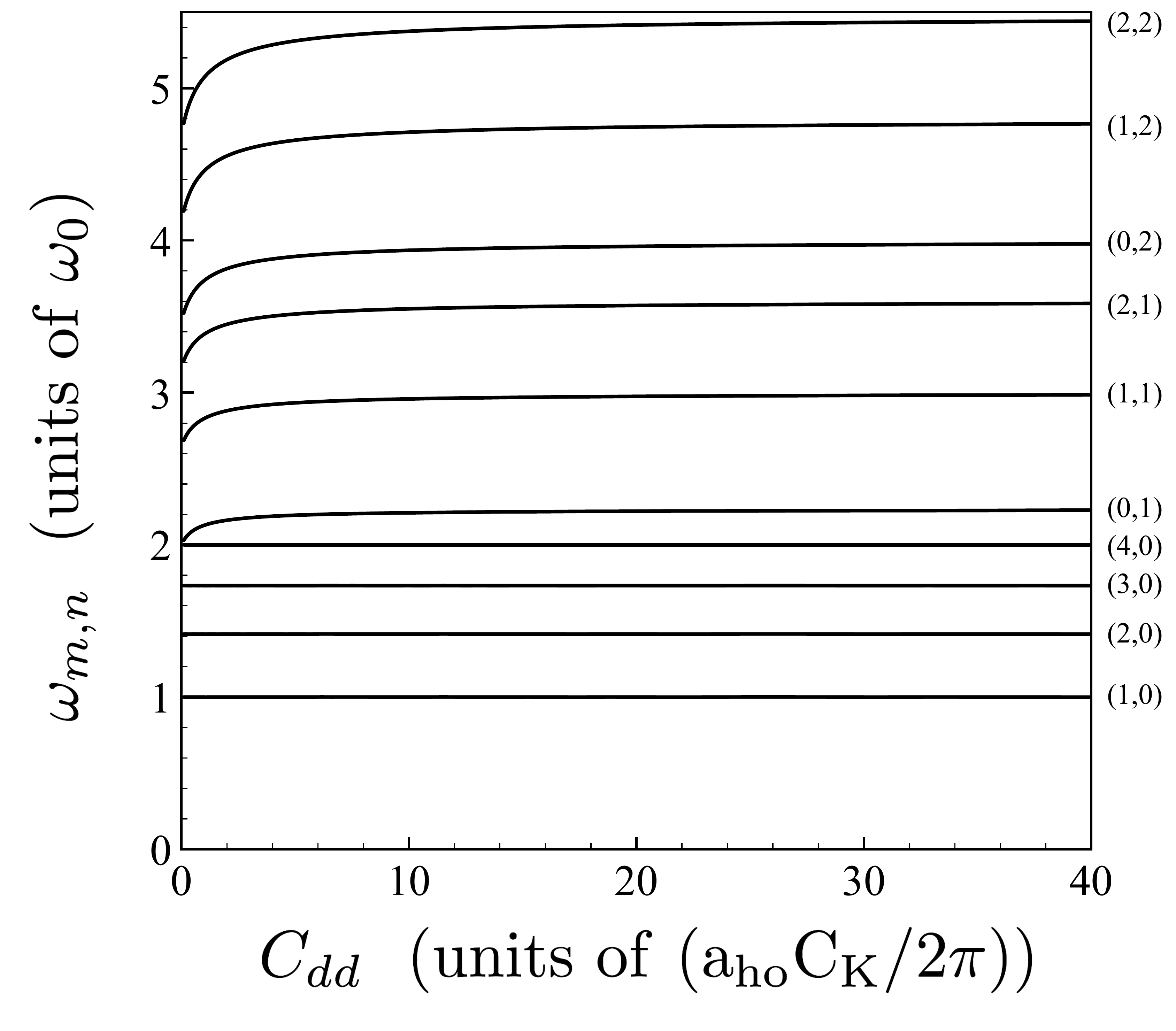}}
\caption{As in Fig.~\ref{fig1}, but  for $N=10^5$ atoms. }
\label{fig4}
\end{figure}

In Fig.~\ref{fig1}, we present a subset of the collective mode spectra corresponding to  $N=10^3$ atoms, with ${C}_{dd}\in [0,40]$ defined in Eq.~\eqref{dimless}.
The value ${C}_{dd}=0$ is only of academic interest as
it corresponds to the noninteracting limit, and the HD regime cannot be reached for any number of particles. The various collective mode frequencies $\omega_{m,n}$ are labeled by $m$, the angular momentum quantum number and $n$, the radial node index.  
It is clear from Fig.~\ref{fig1} that the $n = 0$ modes exhibit a behaviour as a function of $\bCdd$ which is different from the $n\neq 0$ modes.  In particular, the modes for $n\neq 0$
shift up in frequency as ${C}_{dd}$  is increased, eventually saturating to a constant value.  In contrast, the $\omega_{m,0}$ modes appear to be independent of the interaction strength, $C_{dd}$,
on the scale of the plot.   


Apart from the $\omega_{1,0}$ mode, the $n=0$ modes do in fact exhibit a weak dependence on $N$ and $C_{dd}$. This can be seen more clearly in Fig.~\ref{fig2} which shows the dependence of these modes on $C_{dd}$ for $N = 10^2$ and $10^5$. The behaviour of the $\omega_{1,0}$ mode is special since it is governed by the generalized Kohn theorem~\cite{brey}
which states that, for harmonic confinement,
the lowest lying dipolar mode frequency is exactly equal to the trap frequency, independent of the interactions or particle number.  It is therefore noteworthy that the TFvW theory captures this important property for all $N$ and for arbitrary interaction strengths.  On the other hand, the frequencies, $\omega_{m,0}$, of the $m >1$ modes are generally {\em not} independent of ${C}_{dd}$ and exhibit a softening with increasing interaction strength and eventually saturate to a constant value as $C_{dd} \to \infty$. This behaviour is most evident for $N=10^2$, but as Fig.~\ref{fig1} indicates, by $N=10^3$ the $n=0$ modes are virtually flat. In Fig.~\ref{fig4} we show results for $N= 10^5$; by this point, the flat dispersion of the $n=0$ modes is well-established and the $C_{dd}$-dependence of {\it all} the modes has reached a limiting behaviour which corresponds to the TF limit ($\lambda_{\rm vW} = 0$).


Of all the $n\ne 0$ modes, the $m=0,\, n=1$ ``breathing mode'' is of particular interest. The frequency of this mode, $\omega_{0,1}$, starts out at $2 \omega_0$ and
eventually saturates to $\sqrt{5} \omega_0$ as $C_{dd}$ is increased.  This behaviour is significant, since it is known that for a harmonically trapped quantum gas, with a two-body interaction obeying the scaling relation
$V(\beta \br) = \beta^{-2} V(\br)$, there is an underlying $SO(2,1)$ symmetry which ensures the existence of a breathing oscillation at exactly $2\omega_0$, independent of the details of the interaction~\cite{gao,hoffman,pit}.
When we turn off interactions, the TFvW theory has a local kinetic energy density proportional to $n^2(\br)$ which is of the same form as a contact interaction at the mean-field level.
Therefore, the presence of the breathing mode at $2\omega_0$ with $C_{dd}\to 0$ is entirely expected.  However, when $C_{dd}$ differs from zero, its $n^{5/2}(\br)$ contribution to the local energy
density breaks the scale invariance and, as a result, the breathing mode is no longer pinned to $2\omega_0$.

%

\section{Analytical results}\label{analytical}
In this section, we present analytical calculations for the collective modes, in both the weakly ($\bCdd \to 0$) and strongly ($\bCdd \to \infty$) interacting regimes, within the TF approximation (TFA), {\it viz.,}
 $N\gg 1$.  Of particular interest will be an examination of the $\omega_{m,0}$ modes ($m>1)$, which in Sec.~\ref{numerics}, were numerically found to be almost independent of the interaction strength, ${C}_{dd}$, for $N\gtrsim 10^3$.   

The TFvW theory  is reduced to the TFA by simply setting the vW parameter $\lambda_{\rm vW}$ in Eq.~\eqref{SE_TFvW} to zero.  The TFA equilibrium density profile ($n_0(r)=\psi_0^2(r)$) is then determined by
 \begin{equation}\label{TFAden}
 2\pi\psi_0^{2} + \bCdd\psi_0^{3} + \frac{r^{2}}{2} = \mu~.
\end{equation}
While this cubic equation can be solved in closed form for arbitrary $\Cdd$, it is difficult to make analytic progress with this form of the equilibrium wave function.

However, the collective mode spectra can be obtained analytically in the two limiting cases $C_{dd}\to 0$  and $C_{dd}\to \infty$. In these limits, the equilibrium wave function takes the form
\beq\label{psi_general}
\psi_0(r) = \left(\frac{R^2-r^2}{2 C_{\alpha}}\right)^{1/\alpha}~,
\end{equation}
where $\alpha = 2$ for $C_{dd}\to 0$ ($C_2 = C_K$) and  $\alpha = 3$ for $C_{dd}\to \infty$ ($C_3 = C_{dd}$). The constant $R$ is the TF radius, which is fixed by normalizing the density to the correct number of particles, $N$, {\it viz.,}
\beq\label{TFrad}
R =  \left(\frac{2^{(2-\alpha)/\alpha} (2 {\alpha}+4) N C_\alpha^{2/\alpha}}{{\alpha}\pi}\right)^{\alpha/(4+2\alpha)}~.
\eeq
One can view Eq.~\eqref{psi_general} as arising from a polytropic {\em local} energy density of the form 
\beq\label{eloc_gen}
\varepsilon_{\rm loc}[n] =C_{\alpha} \frac{2}{2+\alpha} n^{(2+\alpha)/2}~,
\eeq
with an effective potential given by
\beq\label{veff_2}
v_{\rm eff}^0(r) = \left. \frac{d \varepsilon_{\rm loc}[n]}{dn}\right|_{n=n_0} + \frac{1}{2} r^2 = C_\alpha  n_0(\br)^{\alpha/2}+ \frac{1}{2} r^2~.
\eeq
In the TF limit, the hydrodynamic equation for the density fluctuation, Eq.~\eqref{dyn1},  reads
\bea\label{dyn1_2}
-\omega^2 \delta n(\br) -\nabla \cdot [n_0(\br) \nabla \delta v_{\rm eff}(\br)] &=& 0
\eea
or
\bea\label{dyn1_3}
-\omega^2 \delta n(\br) - \nabla n_0(\br)\cdot \nabla \delta v_{\rm eff}(\br) - n_0(\br) \nabla^2 \delta v_{\rm eff}(\br) &=&0~.
\eea
Eq.~\eqref{eloc_gen} gives for the fluctuating effective potential
\beq\label{fluc_veff_2}
\delta v_{\rm veff}(\br) = \alpha C_{\alpha} [\psi_0(r)]^{\alpha-1} \delta \psi (\br)~.
\eeq

To obtain a solution of Eq.~\eqref{dyn1_3}, we use the wave function fluctuation as the dependent variable, which we write as
$\delta \psi(\br) = [\psi_0(r)]^{1-\alpha} y(r) r^m e^{i m \theta}$. With the change of variable $x=r^2/R^2$,
 Eq.~\eqref{dyn1_3} takes the form of the hypergeometric differential equation~\cite{grad}
 \beq\label{mainODE}
 x(1-x)y''(x) +(c-(a+b+1)x) y'(x) -ab y(x) = 0~,
 \eeq
where primes denote differentiation with respect to the argument, and we have identified
\begin{eqnarray}
a &=& \frac{2+ m\alpha - \sqrt{4+m^2\alpha^2+4\alpha\omega^2}}{2\alpha}\\
b &=& \frac{2+ m\alpha + \sqrt{4+m^2\alpha^2+4\alpha\omega^2}}{2\alpha}\\
c &=& m+1~.
\end{eqnarray}
Note that $C_{\alpha}$ does not appear anywhere in Eq.~\eqref{mainODE}, implying that the mode frequencies, $\omega$, are independent of $C_{\alpha}$  for a polytropic energy density.

Equation~\eqref{mainODE} has two linearly
independent solutions. The appropriate solution is determined by the requirement that the {\em velocity field} be finite
at both $x=0$ and $x=1$. This is equivalent to demanding that the density fluctuation be regular at $x=0$, and that the outgoing current density vanish at $x=1$. 
It is straightforward to show from Eq.~\eqref{lin_mom} that the velocity field is given by
\begin{equation}\label{lin_mom_2}
\begin{split}
\vec{v} 
&\propto  e^{i m \theta} x^{\frac{1}{2}(m-1)}\left[2 x y'(x)\mathbf{\hat{r}}+my(x)(\mathbf{\hat{r}}+i\mathbf{\hat{\theta}})\right]~.
\end{split}
\end{equation}
Equation~\eqref{lin_mom_2} is finite at both $x=0$ and $x=1$ if and only if the function, $y(x)$, and its derivative, $y'(x)$, are regular at both $x=0$ and $x=1$.  In order to satisfy this
condition, the correct solution to Eq.~\eqref{mainODE} is
\beq\label{y}
y(x) ={_2F_1}[a,b,c;x]~,
\eeq
where we must choose $a=-n$ (or equivalently, $b=-n$), $n$ a non-negative integer, in order that the power series expansion of the hypergeometric function, $y(x) =~ _2F_1[a,b,c;x],$ terminate. It follows immediately that the mode frequencies are given by
\beq\label{omega_alpha}
\omega^{(\alpha)}_{m,n} = \sqrt{\alpha n^2 + \alpha m n + 2n + m}~.
\eeq
As stated earlier, the discrete mode spectrum is indeed independent of $C_{\alpha}$ but depends on the polytropic index $\alpha$.  For $\alpha=2$, we are in the $C_{dd}\to 0$ limit, and we obtain
\beq\label{omega_2}
\omega^{(2)}_{m,n} = \sqrt{2 n^2 + 2 m n + 2n + m}~,
\eeq
while for $\alpha=3$, we approach the $C_{dd} \to \infty$ limit, for which the mode spectrum is given by
\beq\label{omega_3}
\omega^{(3)}_{m,n} = \sqrt{3 n^2 + 3 m n + 2n + m}~.
\eeq
Equations~\eqref{omega_2} and \eqref{omega_3} agree perfectly with the numerical TFvW mode spectra for large $N$ in the appropriate limits.  Since $\omega^{(3)}_{m,n} > \omega^{(2)}_{m,n}$, we can now also qualitatively 
understand the reason why
the frequencies of all of the $n \neq 0$ modes are shifted up in frequency as $C_{dd}$ is increased from zero to infinity.


Finally, the density fluctuations are given by
\beq\label{den_fluc}
\delta n_{m,n}(\br) \propto [\psi_0(r)]^{2-\alpha}{_2F_1}[-n,2/\alpha + m + n,m+1;r^2/R^2] r^m e^{i m \theta}~.
\eeq
Owing to the vanishing of $\psi_0(r)$ at the edge of the cloud in the TFA, this result shows that the density fluctuation is regular at $r=R$ for $\alpha \leq 2$, but {\em irregular} for $\alpha > 2$.  Thus the requirement that the density fluctuation be regular at $r=R$ is not always the correct boundary condition to impose in order
to obtain the discrete mode spectrum, Eq.~\eqref{omega_alpha}.  It is also worth  noting that, for
$\lambda_{\rm vW} \neq 0$, the numerically-obtained mode densities do not exhibit any singular behaviour.  However, in the $N\to \infty$ limit they asymptotically approach the TFA solutions.

\subsection{Nodeless excitations}
We observe from Eq.~\eqref{omega_alpha} that the $n=0$ mode frequencies are given by 
$\omega^{(\alpha)}_{m,0}=\sqrt{m}$ for arbitrary values of the polytropic index $\alpha$. However, this does not account for the flat dispersion of the $n=0$ modes found numerically in the large $N$ limit (see dashed curves in Fig.~\ref{fig2}) since the local energy density is not of the polytropic form for arbitrary values of the interaction strength $C_{dd}$. Although the mode frequencies {\em must} be pinned to $\sqrt{m}$ in the weakly and strongly interacting limits, they in principle could exhibit some dispersion as a function of $C_{dd}$.  The flat dispersion of the $n=0$ modes for $N \gg 1$ and {\em arbitrary} $C_{dd}$ is not obvious, and merits some additional discussion.

When $n=0$, $~_2F_1[0,2/\alpha + m ,m+1;r^2/R^2] =1 $, $\forall m \in {\mathbb Z}^+$ and $\forall x \in [0,1]$, and it follows  from Eq.~\eqref{den_fluc} that
\beq\label{den_fluc_2}
\delta n_{m,0}(\br) = A \frac{d n_0(r)}{dr} r^{m-1} e^{i m \theta}~,~~~~(m \geq 1)~,
\eeq
where $A$ is a constant.  The fact that the density fluctuation has this form follows from the assumption of a polytropic local
energy density, {\it viz.,} Eq.~\eqref{eloc_gen}, but its validity is in fact more general.
Assuming that Eq.~\eqref{den_fluc_2} is valid for an {\em arbitrary} local energy density,
the fluctuating effective potential is given by
\bea\label{fveff}
\delta v_{\rm eff}(r,\theta) &=& \left.\frac{d^2\varepsilon_{\rm loc}[n]}{dn^2}\right|_{n=n_0} \delta n_{m,0} \nonumber\\
&=& A r^{m-1} e^{i m \theta} \frac{d}{dr}\left(\left.\frac{d \varepsilon_{\rm loc}}{dn}\right|_{n=n_0}\right)\nonumber \\
&=&  -A r^m e^{i m \theta}~,
\eea
where in going from the second to third line in Eq.~\eqref{fveff} we have made use of the fact that in the TFA with harmonic confinement, the spatial density is defined by
\beq\label{equilden}
\left. \frac{d \varepsilon_{\rm loc}[n]}{dn}\right|_{n=n_0} + \frac{1}{2}r^2=\mu~.
\eeq
Since $r^m e^{i m \theta}$ is a solution to Laplace's equation in 2D, it follows that $\nabla^2 \delta v_{\rm eff} = 0$ and that
\beq
\nabla\cdot (n_0 \nabla \delta v_{\rm eff}) = - m \delta n_{m,0}~.
\eeq
Equation~\eqref{dyn1_3} thus reduces to
\beq
-\omega_{m,0}^2 \delta n_{m,0} + m \delta n_{m,0} = 0~,
\eeq
which immediately yields 
\beq\label{w_m}
\omega_{m,0} = \sqrt{m}~.
\eeq
 We have therefore demonstrated that the $n=0$ density fluctuation in Eq.~\eqref{den_fluc_2} is indeed a solution of the TF hydrodynamic equations with a mode
 spectrum which is {\it independent} of the 
 explicit form of the local energy density $\varepsilon_{\rm loc}[n]$. This conclusion was arrived at earlier using a different approach~\cite{boettcher,Combescot02}.
 A similar result was also found for a two-component Fermi gas interacting {\it via} an $s$-wave contact interaction  by Amoruso {\it et al.}~\cite{amoruso}. However, in this latter work,
 the fact that the nodeless density fluctuations have a frequency $\omega_{m,0}=\sqrt{m}$ was only established in the weak coupling limit.
 

\begin{figure}[ht]
\centering \scalebox{0.7}
{\includegraphics{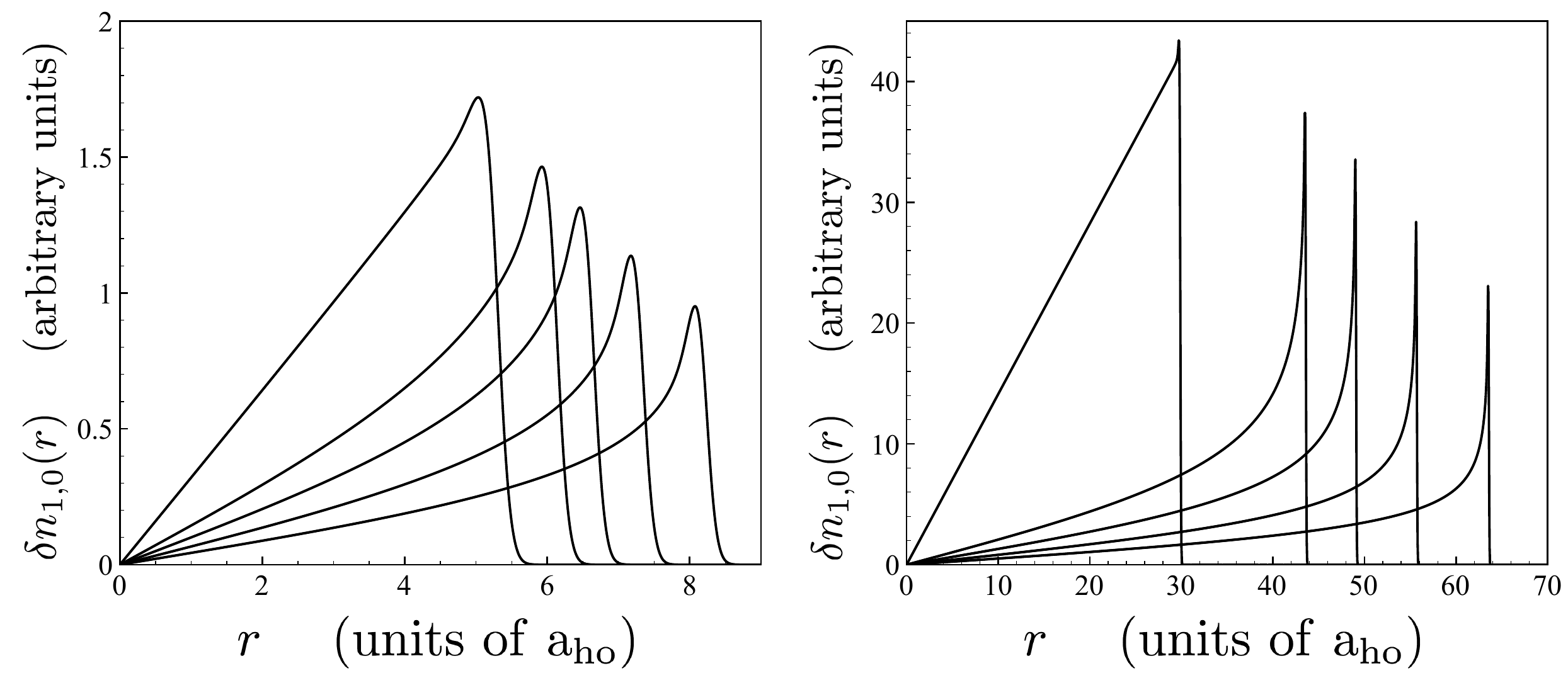}}
\caption{Density fluctuations corresponding to the $\omega_{1,0}$ dipolar (Kohn) mode.  The left panel corresponds to $N=10^2$ particles and the right panel to $N=10^5$ particles.  Within each panel, from left to right, we have $\bCdd = 0,5,10,20,40$ respectively. }
\label{fig5}
\end{figure}

\begin{figure}[ht]
\centering \scalebox{0.7}
{\includegraphics{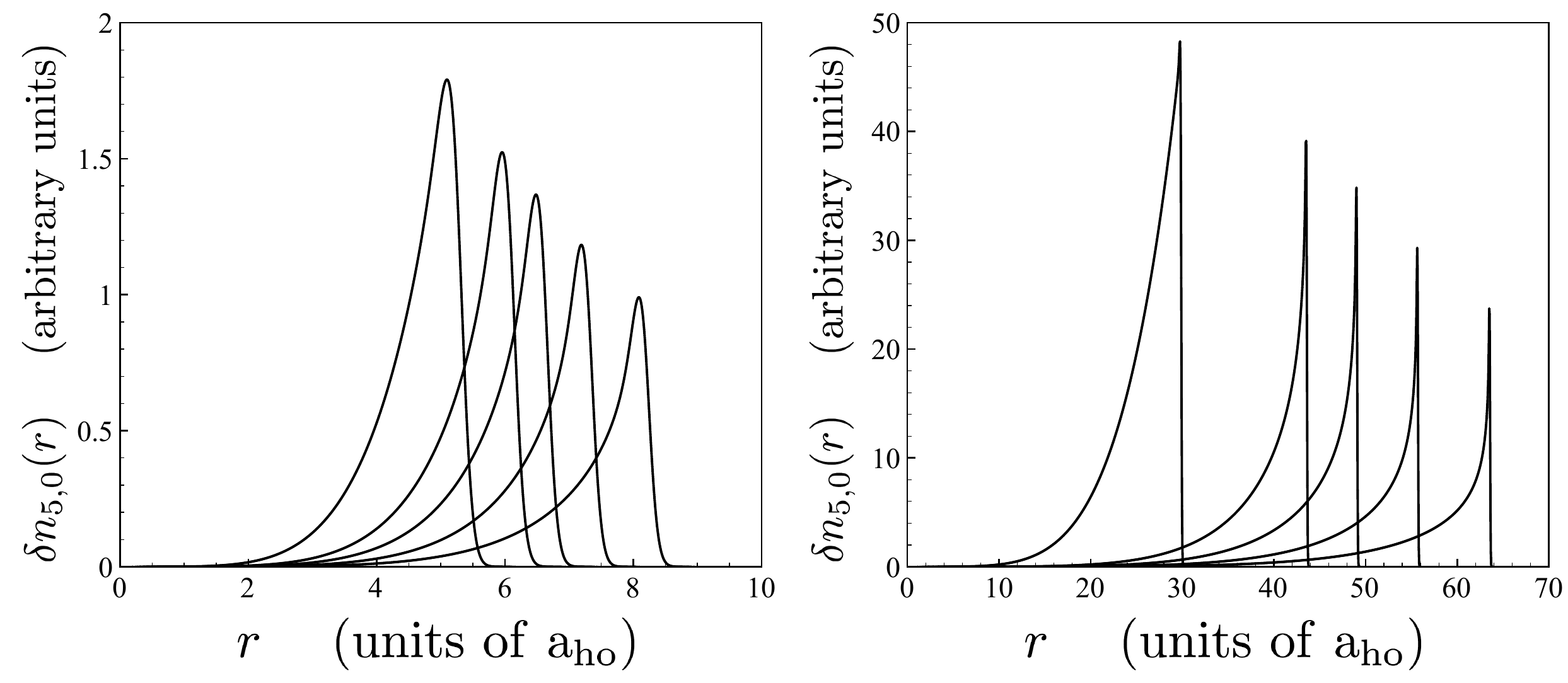}}
\caption{As in Fig.~\ref{fig5} but for the $\omega_{5,0}$ mode.}
\label{fig6}
\end{figure}

In the TFvW theory, the nodeless excitations are not precisely of the form given in Eq.~(\ref{den_fluc_2}), except for $m=1$. According to the generalized Kohn theorem, the lowest dipolar excitation corresponds to a rigid oscillation of the equilibrium density. Thus, for small oscillation amplitudes, the density fluctuation is proportional to the radial derivative of the equilibrium density.
In Fig.~\ref{fig5} we display the density fluctuations, $\delta n_{1,0}(r)$,  for $N=10^2$ (left panel) and $N=10^5$ (right panel) for a variety of interaction
strengths, $\bCdd$. In all cases, the density fluctuations are found to be proportional to $dn_0(r)/dr$ to within numerical accuracy. For $N=10^2$, the smooth nature of the density fluctuation is apparent, but less so for $N=10^5$ where the density fluctuations are approaching the TFA limit. However, for the latter, the sharp feature at the edge of the cloud is the effect of the vW term and would be absent in the strictly TFA calculation.  For example, $\delta n_{1,0}(r) \propto r$ for $C_{dd} = 0$ in the TFA.
Figure~\ref{fig6} illustrates the density
fluctuation, $\delta n_{5,0}$, for the $\omega_{5,0}$ mode for $N=10^2$ (left panel) and $N=10^5$ (right panel).  Similar curves are found for all of the $m>1$ excitations; with increasing $m$, the factor  $r^{m-1}$ leads to an increased localization of the density fluctuation to the edge of the cloud.
\section{Closing Remarks}\label{closing}

We have applied the TFvW hydrodynamic theory to examine the $T=0$ collective modes of a harmonically confined, spin-polarized 2D dFG with purely repulsive interactions.  Our numerical analysis reveals a
rich mode spectrum, with the surface modes ($n=0$) being of particular interest owing to their independence of the interaction strength, even for relatively few particles.  We have argued quite generally that this behaviour
arises from the {\em local} nature of the energy density of the 2D dFG with repulsive interactions, and will not be found in the 1D or 3D dFG, where the interaction is nonlocal.  While the bulk modes ($n\neq 0$) are sensitive to the interaction
strength, for $N \gg 1$, they quickly saturate to a constant value, given by Eq.~\eqref{omega_3}.  We have also provided a detailed
analytical analysis of the system in the TFA, {\it viz.,} $N\gg 1$, which has provided insight into the numerical results found in the full TFvW theory.  

We anticipate that future experiments on $^{161}{\rm Dy}$ in the quasi-2D limit will be faithful realizations of the system discussed in this paper, and be
able to examine the collective excitations  with purely repulsive interactions.  Given that current experiments can readily excite the quadrupole
($\omega_{2,0}$) and breathing modes ($\omega_{0,1}$) of trapped quantum gases, it will be of interest to see if our theoretical predictions for the surface and breathing mode are verified experimentally.  Specifically, in the case of $^{161}{\rm Dy}$, ${\bar C}_{dd} = 2.55$, and 
for $N \sim 10^5$ atoms ({\it i.e.,} well into the HD regime)
we predict the quadrupole mode to have a frequency of $\sqrt{2}\omega_0$, while the breathing mode should have its frequency shifted up by approximately  $6\%$ from its scale-invariant value, $2\omega_0$.

 Finally, a worthwhile extension of this work would be to examine both the equilibrium state and the collective excitations when the dipoles are oriented at some angle relative to the $z$-axis, in which case the
interaction will be nonlocal, and anisotropic.  The inclusion of finite-temperature effects in the formalism would also be of some interest.

\acknowledgements
This work was supported by grants from the Natural Sciences and Engineering Research Council of Canada (NSERC). J. Towers would like to thank Prof. D. A. W. Hutchinson for
valuable discussions, and financial support while completing this work.

\end{document}